\documentclass[journal]{vgtc}                     

\onlineid{1160}

\vgtccategory{Research}

\vgtcpapertype{interaction/representation}

\title{InkSight: Leveraging Sketch Interaction for Documenting Chart Findings in Computational Notebooks}

\author{%
  Yanna Lin,
  Haotian Li, Leni Yang, Aoyu Wu, and 
  Huamin Qu
}
\authorfooter{
\item Yanna Lin, Haotian Li, Leni Yang, Huamin Qu are with the Hong Kong University
of Science and Technology.\\
E-mail: \{ylindg, haotian.li, lyangbb\}@connect.ust.hk,  huamin@cse.ust.hk

\item Aoyu Wu is with the Harvard University. \\
E-mail: aoyuwu@seas.harvard.edu

\item Leni Yang is the corresponding author.
}

\abstract{%
Computational notebooks have become increasingly popular for exploratory data analysis due to their ability to support data exploration and explanation within a single document.
Effective documentation for explaining chart findings during the exploration process is essential as it helps recall and share data analysis.
However, documenting chart findings remains a challenge due to its time-consuming and tedious nature. 
While existing automatic methods alleviate some of the burden on users, they often fail to cater to users' specific interests. 
In response to these limitations, we present \system{}, a mixed-initiative computational notebook plugin that generates finding documentation based on the user's intent.
\system{} allows users to express their intent in specific data subsets through sketching atop visualizations intuitively.
To facilitate this, we designed two types of sketches, \ie~open-path and closed-path sketch.
Upon receiving a user's sketch, \system{} identifies the sketch type and corresponding selected data items. 
Subsequently, it filters data fact types based on the sketch and selected data items  before employing existing automatic data fact recommendation algorithms to infer data facts. 
Using large language models (GPT-3.5), \system{} converts data facts into effective natural language documentation.
Users can conveniently fine-tune the generated documentation within \system{}. 
A user study with 12 participants demonstrated the usability and effectiveness of \system{} in expressing user intent and facilitating chart finding documentation. 
}


\teaser{
  \centering
  \includegraphics[width=\linewidth]{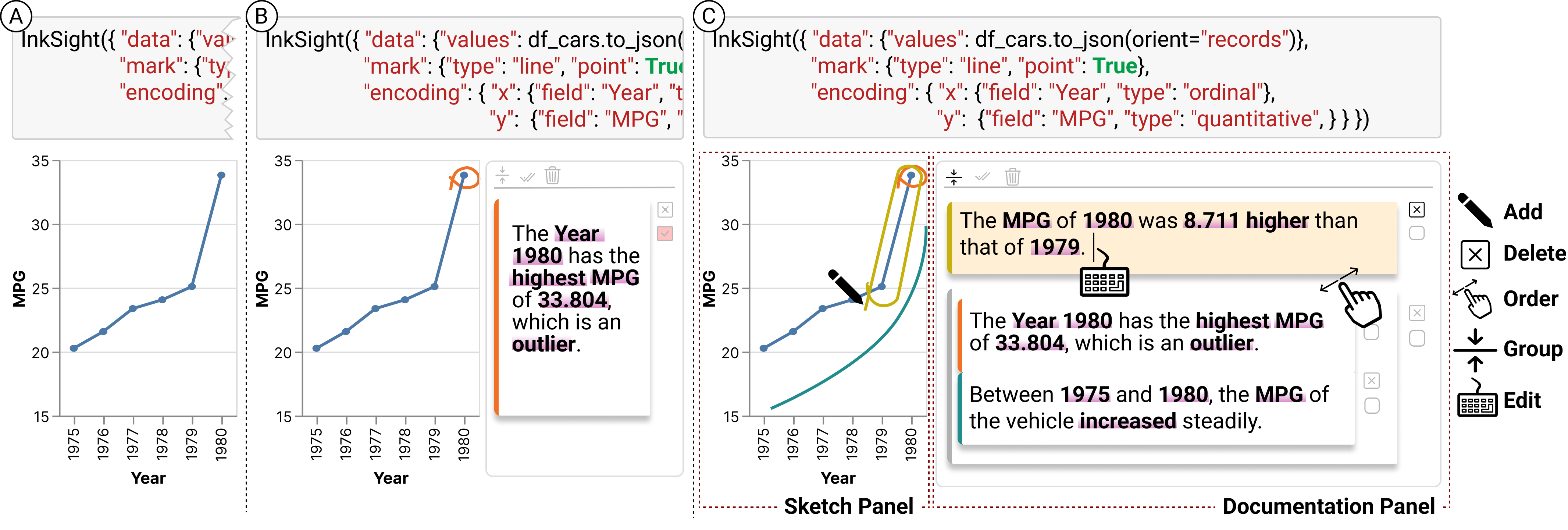}
  \caption{This figure illustrates the process of using \system{} to document findings in a computational notebook. (A) displays the chart created by the user for data analysis. (B) demonstrates that when the user sketches atop the chart to identify areas of interest, \system{} automatically generates corresponding documentation on the right. (C) reveals how users can add more sketches and refine the documentations by performing interactions such as deleting, ordering, grouping, and editing.
  }
  \label{fig:teaser}
  \vspace{-1em}
}

\graphicspath{{figs/}{figures/}{pictures/}{images/}{./}} 

\usepackage{tabu}                      
\usepackage{booktabs}                  
\usepackage{lipsum}                    
\usepackage{mwe}    

\usepackage{tabularx}

\usepackage{mathptmx}                  
\newcommand{\ie}{{i.e.,}\xspace}
\newcommand{\eg}{{e.g.,}\xspace}

\newcommand{\yanna}[1]{\textcolor{black}{#1}}
\newcommand{\haotian}[1]{\textcolor{black}{#1}}
\newcommand{\cl}[1]{\marginpar{  }}
\newcommand{\cll}[1]{\marginpar{\hfill}}

\newcommand{\system}{InkSight}

\begin{document}



\firstsection{Introduction}

\maketitle

Computational notebooks, such as Jupyter~\cite{jupyter} and RStudio~\cite{rstudio}, are increasingly used for iterative data exploration due to their power of combining code, visualizations, and text in a single document.
Yet, there is a gap between the exploration process and explaining the notebook for recalling, sharing, collaboration, and reproducibility of data analysis~\cite{wang2022documentation, rule2018exploration}. 
An explanation should consider both the code and analysis results in which data visualizations are commonly seen~\cite{rule2018exploration}.
\cl{R3.1}
\yanna{However, documenting chart findings is one of the pain points in computational notebooks and remains challenging \cite{chattopadhyay2020s, li2023notable, wang2022documentation}.
Drafting documentation from scratch manually can be a time-consuming and tedious process, causing some data scientists to disregard it for fear of interrupting their analysis flow \cite{rule2018exploration}.
Previous research found that data analysts strongly desire assistance in documenting analysis results~\cite{li2023notable}.}

Recently, documenting findings in computational notebooks has drawn the attention of researchers.
Wang et al.~\cite{wang2022documentation} conducted interview studies and found that automatic methods are needed to reduce \haotian{the burden of documentation}. 
\cl{R4.8}
\yanna{They further developed Themisto, \haotian{which facilitates documentation in a mix-initiative way.
Though it suggests code documentation automatically, it only} provides a start of a sentence as a prompt-based approach to encourage users to document findings, leaving users to complete it on their own.}
\cl{R2.12}
\yanna{To alleviate users' burden, Notable~\cite{li2023notable} generates documentation of chart findings automatically by adopting a data fact recommendation algorithm.
However, it fails to allow users to specify interest in specific data subsets, forcing users to manually document their findings from scratch when the automated documentation deviates from their focus.}
To fill this gap, 
our work aims to provide a tool that allows users to specify their intent for automatically documenting chart analysis results in computational notebooks.


The documentation process should have little workload and should be well integrated into the exploration process.
With this goal in mind, we apply a sketch-based interaction for users to indicate data items of their interest.
We decided upon this design mainly for two reasons:
(1) It has been found that data analysts prefer sketching over keyboard input to maintain their mental flow while externalizing their thoughts during data analysis~\cite{kim2019inking}.
(2) Sketching allows for a more accurate and flexible representation of users' intentions; for instance, it allows analysts to draw any shape to indicate their areas of interest within a chart and captures the nuances of users' drawings.

We present \system{}, a mixed-initiative and on-the-fly computational notebook plugin.
The tool supports users to indicate data items of their interests through easy sketching on charts for document findings.
As shown in \Cref{fig:teaser}, when users create a chart for data analysis (\Cref{fig:teaser} (A)), they can sketch their area of interest directly atop the chart, prompting \system{} to automatically generate the documentation for the related data facts (\Cref{fig:teaser} (B)).
The tool allows users to add more intent and refine the generated documentation through editing, deleting, grouping, and ordering (\Cref{fig:teaser} (C)).
To facilitate the intuitive expression of user intent,
we designed two types of sketches: open-path and closed-path sketches. 
Upon receiving a user's sketch, \system{} aims to infer the user's intent and generate corresponding documentation. 
Specifically, \system{} first identifies the sketch type and corresponding selected data items. 
Next, \system{} filters data fact types based on data types, the number of selected data items, and the sketch type. 
\system{} then employs existing automatic data fact recommendation algorithms~\cite{ding2019quickinsights, wang2019datashot} to infer data facts. 
Finally, \system{} leverages the natural language generation models~\cite{shi2020calliope, gpt3} to convert data facts into natural language, making the documentation more accessible and effective for users.
To demonstrate the usefulness of \system{},  we conducted a user study with 12 participants.
The results show that sketching is efficient and effective in catching the users' intent, and  \system{} can help document findings.
Finally, we further summarized the lessons we learned and discussed potential future directions.
In summary, our main contributions are as follows:

\begin{itemize}
    \item We provide a convenient and efficient approach for users to express intent by sketching atop the visualizations directly;
    \item We present \system{}, a computational notebook plugin, enabling users to document findings easily;
    \item We conduct a user study to demonstrate the usefulness and effectiveness of \system{}.
\end{itemize}

\section{Related Work}

Our research is related to prior studies on visual data analysis for notebooks, documentation, visualization summarization, and intent expression.

\subsection{Visual Data Analysis in Computational Notebook}
With the popularity of computational notebooks for exploratory data analysis, facilitating visual data analysis in computational notebook environments has become a trending research topic in the visualization and HCI community.
Their focuses range from creating data visualizations to presenting the analysis results.
For example, researchers have developed Jupyter Notebook extensions that generate visualizations from dataframe transformation code~\cite{wu2020b2}, printed dataframes~\cite{lee2021lux}, and data queries~\cite{lee2021lux,chen2022pi2}.
Overall, much effort has been put into releasing users from laborious code writing for data wrangling and visualization during the data exploration process.

Besides exploration, the role of computational notebooks for documenting and sharing analysis results raises another critical issue.
That is, how to facilitate results documentation which could interrupt the analysis flow and be discarded by users as being too demanding\yanna{~\cite{rule2018exploration,wang2022documentation, chattopadhyay2020s}}.
Themisto~\cite{wang2022documentation} applied deep-learning algorithms to generate text for explaining codes in computational notebooks. 
However, for documenting analysis results, it only generates a few beginning words to prompt users to start writing.
More work introduced Human-AI collaboration tools for generating presentation slides from computational notebooks.
For example, NB2Slides~\cite{zheng2022nb2slides} and Slide4N~\cite{wang2023slide4n} automatically distill key messages from cells in computational notebooks and organize them into slides.
Nonetheless,
users still need to manually write their findings from analyzing charts.
Notable~\cite{li2023notable} takes a step further by extracting data facts from the data in user-specified charts.
Yet,
some participants in the user study suggested dissatisfaction with the divergence between automatically generated results and their intentions.
This observation 
necessitates intuitive and efficient interaction to convey user intent as input for automatic algorithms.

We present \system{}, a tool that provides a sketch-based interaction to allow users to suggest data items of their interests through easy sketching atop data visualizations. 
Compared with previous work, \system{} captures user intent at a more fine-grained level without a significant increase in user interaction and workload.
Instead of generating a whole slide presentation, we focused on integrating the documented findings in the same notebook for sharing and future recall.

\subsection{Natural Language and Visualization}
\cl{SR.4, R4.9}
\yanna{Visualizations are frequently coupled with text to convey information effectively with data and tell compelling stories~\cite{yang2021design}.
Some research efforts have been invested in linking text and visual elements within visualization to eliminate the need for shifting users' attention and enhancing comprehension~\cite{latif2022kori, kong2014textchart}.
Diverging from linking text with visualizations, another line of research aims at generating one form of communication from the other.
Much of this research centers on generating visualizations given NL utterances. }
For example, systems such as Text-to-Viz~\cite{cui2019text}, FlowSense~\cite{yu2019flowsense}, GVQA~\cite{song2023gvqa} and NL4DV~\cite{narechania2020nl4dv} apply a semantic parser to convert NL queries into visualization queries.
More recent research like~\cite{wang2022towards} improves the performance by training deep learning models.

Relatively less research has studied the reverse problem, that is, to generate NL descriptions given a visualization.
Researchers have proposed datasets for chart-to-text generation~\cite{kanthara2022chart}.
However, their benchmark results show that state-of-the-art methods still yield unsatisfactory results and often suffer factual errors.
Therefore, many systems first generate factually correct data findings and subsequently generate NL descriptions.
For example, Voder~\cite{Srinivasan2019voder} implements heuristics to extract data insights from tables and generate corresponding NL descriptions and visualizations using templates.
AutoCaption~\cite{liu2020autocaption} extends this workflow to recognize
significant features of the chart images and subsequently generate NL descriptions.
Those techniques are deployed for an array of applications such as iterative data analysis~\cite{Srinivasan2019voder, zhao2021chartstory}, providing text summaries~\cite{liu2020autocaption}, and making charts accessible to individuals with visual impairments~\cite{mishra2022chartvi, obeid2020chart}.
Although those methods capture interesting data findings, they fall short of capturing user intent, which is critical for effective human-computer collaboration.
\cl{SR.2, R2.1}
\yanna{To address this problem, Choi and Jo~\cite{choi2022intentable} introduced Intentable, a mixed-initiative system that allows users to specify their intent by selecting pre-defined intent types and clicking the data items of interest in the chart. 
However, its interaction is limited to clicking, which is inefficient when dealing with a large number of data items of interest. 
For instance, selecting numerous data items in scatterplots, as shown in Figure 6, becomes impractical and time-consuming.
Moreover, it primarily focuses on developing a grammar for intent specification and training a neural network to generate caption sentences.
In contrast, our approach contributes a new interaction design, leveraging sketching as a more natural and expressive method for specifying user intent.}

\subsection{Sketching Interaction in Visual Analysis}
Introduced by Sutherland in the 1960s with the Sketchpad concept~\cite{edward1963sketchpad}, sketching has been widely studied due to its central role in the design process and visual thinking. 
\cl{SR.4, R2.4}
\yanna{In visualizations, sketch-based interaction was first employed for data queries~\cite{ryall2005querylines, siddiqui2020shapesearch, wattenberg2001sketching, holz2009timeseriesquery}}.
For instance, 
ShapeSearch~\cite{siddiqui2020shapesearch} and QuerySketch~\cite{wattenberg2001sketching} allowed users to draw freeform sketches to query time-series data with matching patterns.
Recognizing the benefits of promoting thinking, insight, and inspiration~\cite{tohidi2006getting}, sketch-based interactions have been further employed for data exploration.
Examples include NapkinVis~\cite{chao2010napkinvis}, which enables users to sketch predefined gestures for rapid and effortless visualization creation, and SketchVis~\cite{browne2011sketchvis}, which allows hand-drawn sketch inputs to explore data through simple charts. 
Some research has extended these interactions to pen and touch for creating and manipulating data visualizations~\cite{lee2013sketchstory, walny2012understanding}.

Beyond data exploration, sketching has been identified as a flexible and lightweight way to convey users' high-level intents\yanna{~\cite{chung2022talebrush, eitz2012how, chen2018forte}}. 
For example, TaleBrush~\cite{chung2022talebrush} allows users to generate stories by sketching the protagonist's fortune changes throughout the narrative. 
\cl{SR.4, R2.4}
\yanna{Forte~\cite{chen2018forte} enables users to generate structures for creating 3D fabrication-ready models by sketching the draft plan and loads.}
Kim et al.~\cite{kim2019inking} have observed that users prefer to use digital pens to express their intentions for identifying trends, anomalies, and more.
Given the advantages of sketch-based interactions, such as flexibility, expressiveness, and the ability to convey high-level intentions, \system{} chooses to utilize sketching as a means for users to express their intent.
Our work aims at a new application scenario where sketch interaction will be integrated into the visual data analysis process with computational notebooks of which most interactions are keyboard based.
We believe our study results can complement previous work by providing insights into sketch interactions for visual data analysis in a scenario. 







\section{Design Goal}
Our tool aims at reducing both users' mental and labor effort of documenting chart findings in computational notebooks.
We set ourselves the following design goals. 

\textbf{G1: Offer automatically generated documentation of chart findings.} 
Manually documenting findings can be both time-consuming and tedious, thus some data scientists disregard it~\cite{rule2018exploration}.
Data analysts have expressed a preference for automated approaches that assist in documentation~\cite{wang2022documentation}.
The tool should automatically generate chart findings, reducing the burden of starting the documentation from scratch.

\textbf{G2: Support flexible and effortless user specification of intent.}
Prior research indicates the importance of considering user intent in subsets of data and data fact types to optimize automatic algorithms for creating data visualizations and generating chart findings~\cite{pandey2023medley, lee2021lux,li2023notable, wang2022documentation, shen2021taskvis}.
Critically, the forms of specification affect the efficiency of expressing user intent.
The tool should offer a flexible and effortless way for users to specify their intent.

\textbf{G3: Reduce context switch between exploration and explanation.} 
Context switching between exploration and explanation can be costly~\cite{li2023notable, chen2022crossdata}.
The tool should provide a smooth and natural interaction that minimizes disruption to users' mental flow to reduce users' mental effort required to transition between exploration and explanation tasks.

\textbf{G4: Facilitate iterative refinement of the documentation.} 
Providing users with some control over the generated results can serve as a means of fine-tuning and refining the outcomes to better meet individual user requirements and compensate for the limitations of the automated methods. 
This can improve user satisfaction and increase user engagement with the system~\cite{wang2022documentation}. 
Thus, the tool should facilitate iterative refinement of the generated results conveniently with a suite of interactions.

\textbf{G5: Integrating with existing platform.}
The tool should be developed as a plugin for computational notebooks (\ie~Jupyter Notebook in our case), such that users do not need to be familiar with a new interface.
The chart findings documentation is directly inserted into the notebook as a whole for recalling and sharing.








\section{\system{}}\label{sec:method}

In this section, we first present an overview of \system{} (\Cref{sec: overview}).
Then we introduce the interactive modules and computational modules of \system{} (\Cref{sec:interactive module} and \Cref{sec: computation module}, respectively).

\subsection{System Overview}
\label{sec: overview}
In accordance with the design goals, we developed \system{}, a mixed-initiative Jupyter Notebook plugin (\textbf{G5}) designed to offer a seamless experience for documenting chart findings during data exploration (\textbf{G3}).
In this section, we present the overview of \system{} and elaborate on the details in subsequent sections.
The tool consists of interactive and computational modules.
Interactive modules define the interface and interaction designs.
Computational modules support the functionalities of interactive modules.

As shown in \Cref{fig:teaser} (C), our tool has two interactive modules: the sketch panel and the documentation panel.
For every code cell that creates a data visualization, the two panels are inserted below it. 
In the left sketch panel, users can sketch their intent in subsets of data atop the chart directly (\textbf{G2}).
We pre-defined sketch types tailored to different types of data facts.
For example, users can circle out an outlier point in a scatter plot or draw a line following the trend of a line chart, and the algorithms will generate outlier or trend data facts respectively.
When users explore the chart and identify interesting patterns, they can immediately sketch related areas in the chart to smoothly transition to the task of documenting chart findings (\textbf{G3}).

The right documentation panel displays the generated data facts corresponding to each sketch (\textbf{G1}).
Its design centers around the goal of facilitating the iterative refinement of the documentation (\textbf{G4}).
Users then can refine the documentation by revising, reorganizing, deleting, and grouping multiple data facts.
Furthermore, the sketch panel and documentation panel are linked visually and interactively, inspired by the use of embellishments in Voder to enhance users' understanding~\cite{Srinivasan2019voder}. 
For example, \system{} assigned a color mark to the generated findings with the same color as the corresponding sketch. 
It also enables users to hover over documentation to highlight the corresponding sketch and to click on a sketch to automatically scroll its associated documentation into view. 
The final documentation of chart findings is presented in a hierarchical list structure. 

To support the interactive modules in the front-end user interface, we developed three computational modules (\ie~sketch identification, intent inference, and finding documentation generation) for automatic findings documentation generation (\textbf{G1}). 
Once a user completes a sketch, the sketch identification module interprets it to determine the sketch type and associated data items. 
The intent inference module then infers potential data facts related to these data items and sketch types. 
The finding documentation generation module accepts these data facts and employs a template-based approach to generate descriptive text narratives for each.
It utilizes the advanced language model, GPT-3.5~\cite{gpt3}, to organize and refine all text narratives.
The generated narrative for each sketch is displayed in the documentation panel within the interface.

\begin{figure}[t!]
    \centering
    \includegraphics[width=1\linewidth]{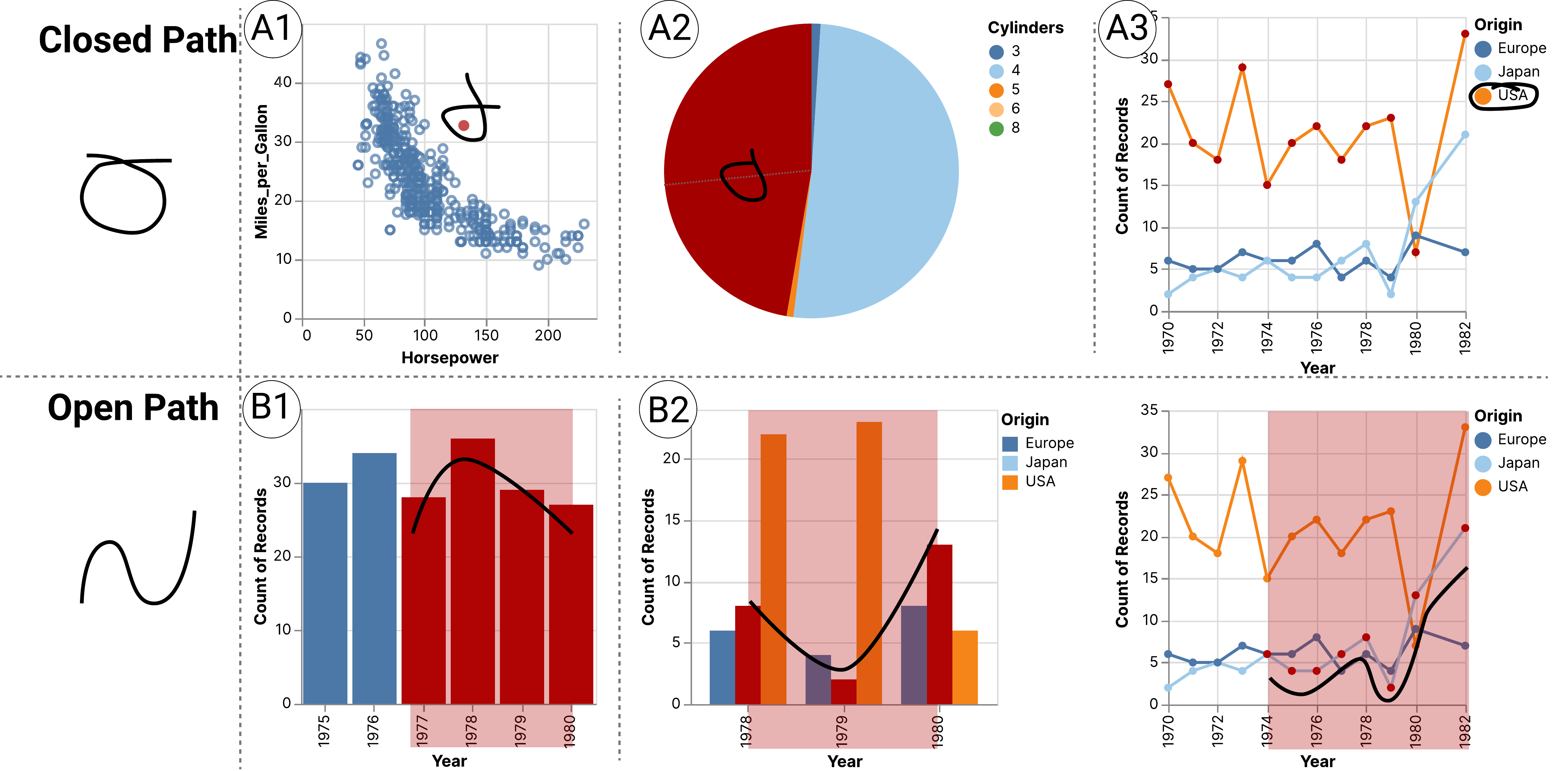}
    \caption{This figure showcases how users can employ both closed-path sketches (A1-A3) and open-path sketches (B1-B2) to express their intent by selecting data subsets of interest. Specifically, the selected data items are highlighted in red, while the selected range features a red background.}
    \label{fig:sketch}
    \vspace{-1em}
\end{figure}

\begin{figure*}[ht]
    \centering
    \includegraphics[width=1\linewidth]{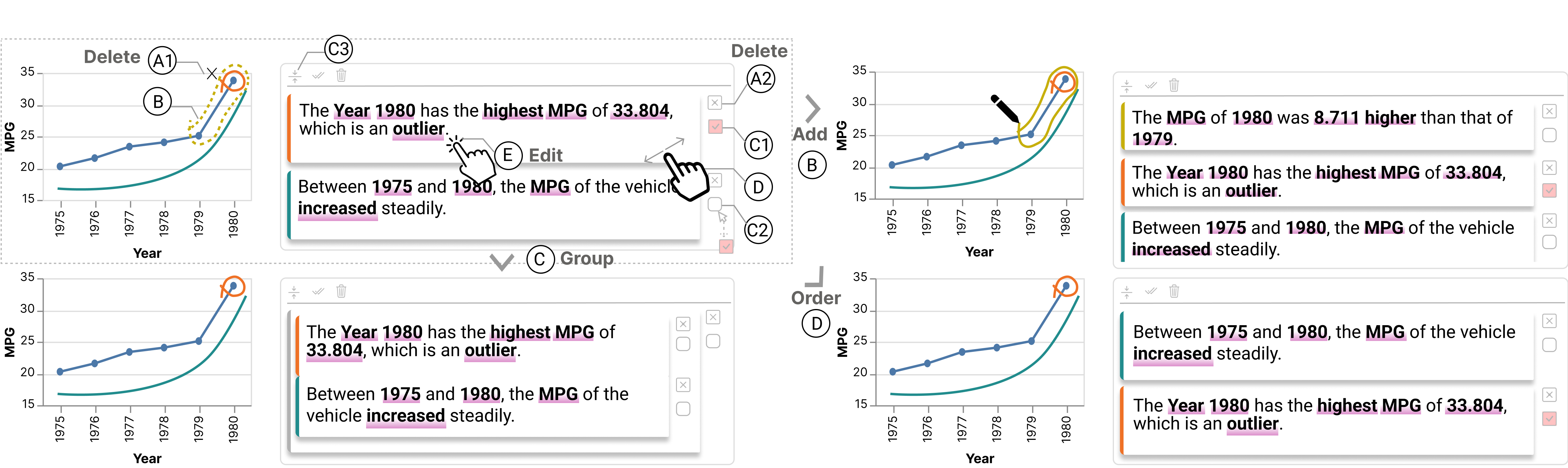}
    \caption{A suite of interactions is provided for refining documentation, illustrating how users can interact with the interface and obtain corresponding results. Specifically, (A1) or (A2) are used to delete the sketch and the corresponding documentation card, (B) denotes drawing a sketch to insert a documentation card, (C) refers to selecting multiple documentation cards for grouping, (D) represents ordering the documentation cards by dragging and dropping, and (E) signifies clicking to edit.}
    \label{fig:interaction}
    \vspace{-1em}
\end{figure*}

\subsection{Interactive Modules}
\label{sec:interactive module}
This section presents two interactive modules in the front-end user interface: 
the sketch panel and the documentation panel.
The sketch panel enables users to express their intents through sketches atop the charts, while the documentation panel provides a suite of interactions to support the convenient refinement of the generated documentation.

\textbf{Sketch Panel.}
As shown in \Cref{fig:teaser} (C), the sketch panel displays the chart (in SVG format) from which users have findings to be documented.
\cl{SR.3, R1.1}
\yanna{Our tool supports basic statistical charts, including bar charts, line charts, pie charts, grouped bar charts, multiple line charts, and scatter plots.}
In the sketch panel, users can sketch directly on top of the visualizations. 
The sketching feature enables users to express their intents in data subsets of interest.
Inspired by~\cite{akers2006cinch}, 
we support two sketch types for expressing intent: closed path and open path.

\Cref{fig:sketch} showcases six examples that illustrate how users can express their intent using open-path sketches (\Cref{fig:sketch} (A1)-(A3)) and closed-path sketches (\Cref{fig:sketch} (B1)-(B2)). 
For the closed path, users encircle data items of interest by ensuring they lie within or intersect the sketch area as shown in A1 and A2.
However, grouped bar charts and multiple line charts involve several data groups each of which could be hard to single out by a close path.
Thus, we enhance users to select a group of interested data items through legends (\Cref{fig:sketch} (A3)) to improve usability.
Circling out a label in the legend represents the selection of all data items that belong to the category.

Regarding the open path, 
users can draw a line to follow each data item of interest in the charts in a loose way. 
To be more specific, 
B1 showcases how users can identify a data range of interest using an open path for simple charts.
Similar to the closed path, the tool offers a convenient way to select a group of data from the same category in multiple line charts and grouped bar charts.
To address this, we enable users to draw a sketch resembling the path of a specific group of data items they find interesting (\Cref{fig:sketch} (B2)).

During drawing, to assist users in checking whether the sketch captures their  data subsets of interest accurately, the sketch panel highlights the selected data items in red (\Cref{fig:sketch} (A1)-(A3) and (B1)-(B2)) or the selected range with a red background (\Cref{fig:sketch} (B1)- (B2)).
To avoid visual clutter, this highlighting effect lasts for one second before disappearing.

\textbf{Documentation Panel.}
The documentation panel on the right is for display and refinement of the generated documentation (\Cref{fig:teaser} (C)).
The documentation panel presents findings in a hierarchical structure consisting of different cards. 
In particular, a card at the lowest level contains findings generated from one sketch.
The card-based visual design is to resemble the code cell-based design in Jupyter Notebook, to ensure a consistent style throughout the interface. 

Once the documentation for each sketch is generated, it is automatically inserted as the first card in the documentation panel, as shown in \Cref{fig:interaction} (B).
Each card is accompanied by two buttons on its right side.
At the top is a deletion button for deleting the documentation card and the corresponding sketch (\Cref{fig:interaction} (A2)).
Alternatively, users can delete a documentation card by clicking on the corresponding sketch and clicking the deletion button that pops up subsequently (\Cref{fig:interaction} (A1)).
At the bottom is a checkbox button for users to select multiple documentation cards and group them together to create a simple hierarchical structure by clicking the grouping button (\Cref{fig:interaction} (C)).
The grouping button is the first button at the top toolbar of the panel \Cref{fig:interaction} (C3)).
To further organize the documentation cards, users can utilize the drag-and-drop operations to reorder the cards within, outside, and between the merged groups and documentation cards (\Cref{fig:interaction} (D)). 
At the top toolbar, three buttons from left to right are for users to group selected cards, group all cards, or delete all cards, respectively. 
Last but not least, users can edit the content of documentation cards by simply clicking on them (\Cref{fig:interaction} (E)).

To enhance the linkage between the documentations and the sketches, our tools provide the following designs.
First, the border of each card is assigned the same color as the corresponding sketch.
Moreover, when users click a sketch on the chart, the corresponding documentation card will scroll into view, thereby facilitating their navigation.
\cll{R4.7}
\yanna{
\haotian{Inspired by a previous study~\cite{yu2021readability}, \system{} highlights key messages in data facts to enhance readability and improve efficiency.}
To highlight the key messages, we first identify all key messages in data facts, including the related data variables' names and values, as well as data patterns revealed by the data facts like increasing or decreasing trends.
Then, utilizing an exact match method, these key messages are located within the sentences and subsequently highlighted for emphasis.}

\begin{table*}[ht]
\caption{The table displays eight data fact types and their corresponding attributes, including the data types involved, the number of data points related,  and the text description template. \textit{C, T, N} indicate the categorical, temporal, and numerical data types, respectively.  \textit{Breakdown} represents the independent variables of a chart, \textit{measure} refers to the dependent variables, \textit{focus} denotes the data points related to the data fact, and others such as \textit{rank} and \textit{corr} are parameters which describe the data fact.}
\begin{tabular}{ccccp{10.5cm}}
\toprule
\textbf{Data Fact Type} & \textbf{Breakdown} & \textbf{Measure} & \textbf{\#Focus}   & \textbf{Template}                                                                                                                                              \\ \midrule
Value               & C/T/N                  & N               & 1                & \{focus\}.                                                                                                                                                     \\ \midrule
Proportion           & C/T                & N                & 1                & The \{focus\} accounts for \{proportion\} of the \{measure\}.                                                                                                  \\ \midrule
Outlier               & C/T                & N                & 1                & The \{focus\} is an outlier in \{measure\}.                                                                                                                    \\ \midrule
Outlier\_scatter     & N                  & N                & 1                & The \{focus\} is an outlier.                                                                                                                                   \\ \midrule
Rank                   & C/T                & N                & 1                & The \{focus\} ranks \{rank\} in \{measure\}.                                                                                                                   \\ \midrule
Difference            & C/T                & N                & 2                & The \{measure\} of \{focus1\} is \{ratio times/difference\} higher than that of \{focus2\}.                                                                    \\ \midrule
Trend          & T                  & N              & \textgreater{}=3 & During \{range\}, the \{measure\} (of the category of \{focus\}) has an \{increasing/decreasing/flat/wavering\} trend over the \{breakdown\} with a slope of \{slope\}. \\ \midrule
Association         & N                 & N              & \textgreater{}=3 & \yanna{\{measure1\} has a \{strong/moderate/weak\} \{positive/negative\} influence to \{measure2\}, with a Pearson correlation as \{corr\}. }  \\                  \bottomrule                                            
\end{tabular}
\label{table:fact}
\vspace{-1em}

\end{table*}

\begin{figure}[h!]
\centering
    \includegraphics[width=1\linewidth]{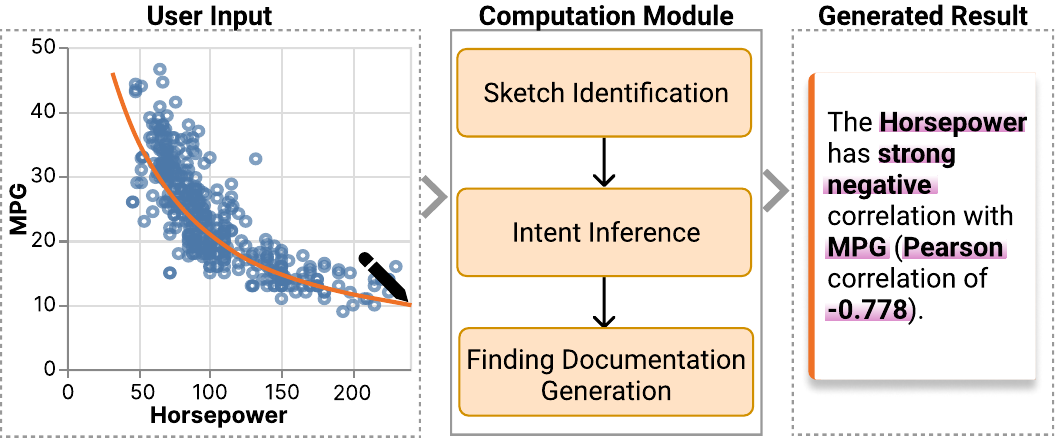}
    \caption{This figure shows the workflow of \system{}. With the user's sketch as input, \system{} contains three computation modules to generate the corresponding documentation, \ie~sketch identification, intent inference, and finding documentation generation.}
    \label{fig:workflow}
    \vspace{-1em}
\end{figure}

\subsection{Computation Modules}
\label{sec: computation module}
This section describes the three computation modules that support the front-end user interface, as depicted in \Cref{fig:workflow}.
First, the sketch identification module takes the Vega-Lite chart specification and the sketch atop the chart as input to identify the sketch types and infer the data items of interest. 
Subsequently, the intent inference module infers a list of potential data facts related to these items. 
Finally, the finding documentation generation module automatically generates and organizes text descriptions for the data facts, resulting in the final documentation.

\textbf{Sketch Identification}.
The sketch identification module aims to determine the type of sketch and extract the data that binds to the sketch.
Once users complete sketching, the sketch identification module first detects the path type (i.e., close or open) and then calculates the data items that bind to the sketch.

For a closed path, the module identifies data items of interest by examining the visual graphical elements included within or intersecting the sketch.
Specifically, if these graphical elements are visual representations of data items, then these data items are considered to be of interest.
Alternatively, if the graphical elements are data labels in the legend, the data items belonging to the selected categories are considered of interest.

For open path, the module identifies data items of interest differently for simple charts (\ie~line chart, bar chart, scatterplot, pie chart) and complex charts (\ie~multiple line chart and group bar chart), which have several data groups from distinct categories.
For simple charts, the module recognizes the selected data range and filters data items within the specified range.
However, in the case of complex charts such as multiple line charts and grouped bar charts, the selected data range may encompass several data groups from distinct categories, necessitating the determination of which data group is of primary interest to the user.
To address this, the module infers that the data group with the highest spatial similarity to the user's sketch is the target group. 
The similarity between each data group and the sketch is evaluated by calculating the average of the shortest distance between each point in the data group and the sketch.
We use the average distance instead of the sum of distances because we observe that different groups may have varying numbers of points due to missing values, within the selected range. 
Consequently, the average distance is employed to mitigate the impact of missing values.
In summary, with the sketch and chart specification as input, the sketch identification module calculates the sketch type and the data items of interest.


\textbf{Intent Inference.}
Given the chart specification and the output of the sketch identification module (including sketch types and data items of interest) as input, the intent inference module generates relevant data facts.
\system{} supports eight data fact types defined by Calliope~\cite{shi2020calliope}, as shown in \Cref{table:fact}. 
\system{} adopts the existing data fact generation algorithm~\cite{ding2019quickinsights, wang2019datashot} to generate data facts.
\cl{SR.3, R2.5, R4.1, R4.5}
\yanna{For example, for the data fact \textit{outlier}, we utilize the Interquartile Range (IQR) method~\cite{IQR} to detect outliers. Outliers are identified as data items with values that are either smaller than $Q1-1.5IQR$ or larger than $Q3+1.5IQR$. The detailed methods are included in the supplementary material. 
}


The module first eliminates meaningless data facts based on data types, the size of the data subset, and sketch types.
For data types, some data fact types can only be from certain types of data (\Cref{table:fact}: column ``Breakdown'' ).
For example, a \textit{trend} must be from a chart that encodes temporal data and numerical data.
Similarly, some data facts have constraints on the minimal size of the data subset (\Cref{table:fact}: column ``\#Focus'').
For example, \textit{difference} compares at least two data items.
Regarding sketch types, we assume that users utilize open-path sketches to express specific interests in trends or associations related to shapes rather than points. 
For closed-path sketches, we apply almost all fact types.
However, if there are multiple selected data items, it is impractical to report all the ranks for each data item in the subset and compare each pair of data items. 
In this case, we only consider the top one and the last one for the \textit{rank} fact and their difference for the \textit{difference} fact.
This distinction between sketch types helps ensure that the inferred data facts are relevant and meaningful.


Next, the module defines the following special cases.
\cl{R2.6, R3.2}
\yanna{First, when a single data item is selected in complex charts comprising multiple groups of data from different categories, inferring users' intent can be challenging.
It is hard to identify whether the users aim to compare the particular data item with others in the same group, with those that have the same independent variable, or with all other data items.
To resolve this, we present all three possibilities to the user, enabling them to select the most appropriate option by deleting the unwanted ones}.
Second, when users draw a closed-path sketch involving multiple items, we assume they are only interested in data items inside the selected data subset without comparing them with data items outside.
In other words, the intent inference module treats the selected data items as a completely separate dataset to calculate the data facts. 

\textbf{Finding Documentation Generation.}
Given the generated data facts from the intent inference module, the finding documentation generation module organizes them into the final generated documentation.
The module initially employs template-based methods, commonly used for natural language generation, to generate text descriptions for each data fact.
\Cref{table:fact} displays the templates employed for each data fact.
Moreover, we add a description of the selected range of data when users select multiple items, such as ``\textit{among the selected $x$ items}'', where $x$ represents the number of selected items.
Subsequently, to improve the coherence and naturalness of the generated results, we leverage an advanced language model, GPT-3.5~\cite{gpt3}, to help organize and refine the data facts.
When there is only one data fact, we ask GPT-3.5 to polish it to be more accurate, concise, and human-like.
\cl{SR.3, \ R1.2, R4.1}
\yanna{In the case of multiple data facts, we have additional requirements.
We instruct GPT-3.5 to synthesize the facts by merging those with similar information and comparing them as instructed in the prompt. 
An example instruction is \textit{``merge those with similar information''}.
Further details about the prompt can be found in the supplementary material.}
\cl{R2.6}
\haotian{The generated documentation can also be edited by the users in the documentation panel. 
In this way, users can revise or remove the data facts that are not aligned with their intent.}


When designing the function of grouping user-selected cards in the documentation panel (\Cref{sec:interactive module}), we also considered merging selected cards into a single sentence.
However, most documentation results already comprise several data facts.
Merging them will generate an excessively long and complex sentence that would be difficult for users to read and comprehend. 
Furthermore, the merged documentation can introduce ambiguity or confusion, as users may struggle to identify the relation between the original documentation and the merged result.
In light of these concerns, we ultimately opted to group multiple documentation cards into a simple hierarchical structure without merging the content. 
This approach preserves clarity and maintains the original information, ensuring that each data fact can be easily understood.





\section{Usage Scenario}

\begin{figure*}[ht]
    \centering
    \includegraphics[width=1\linewidth]{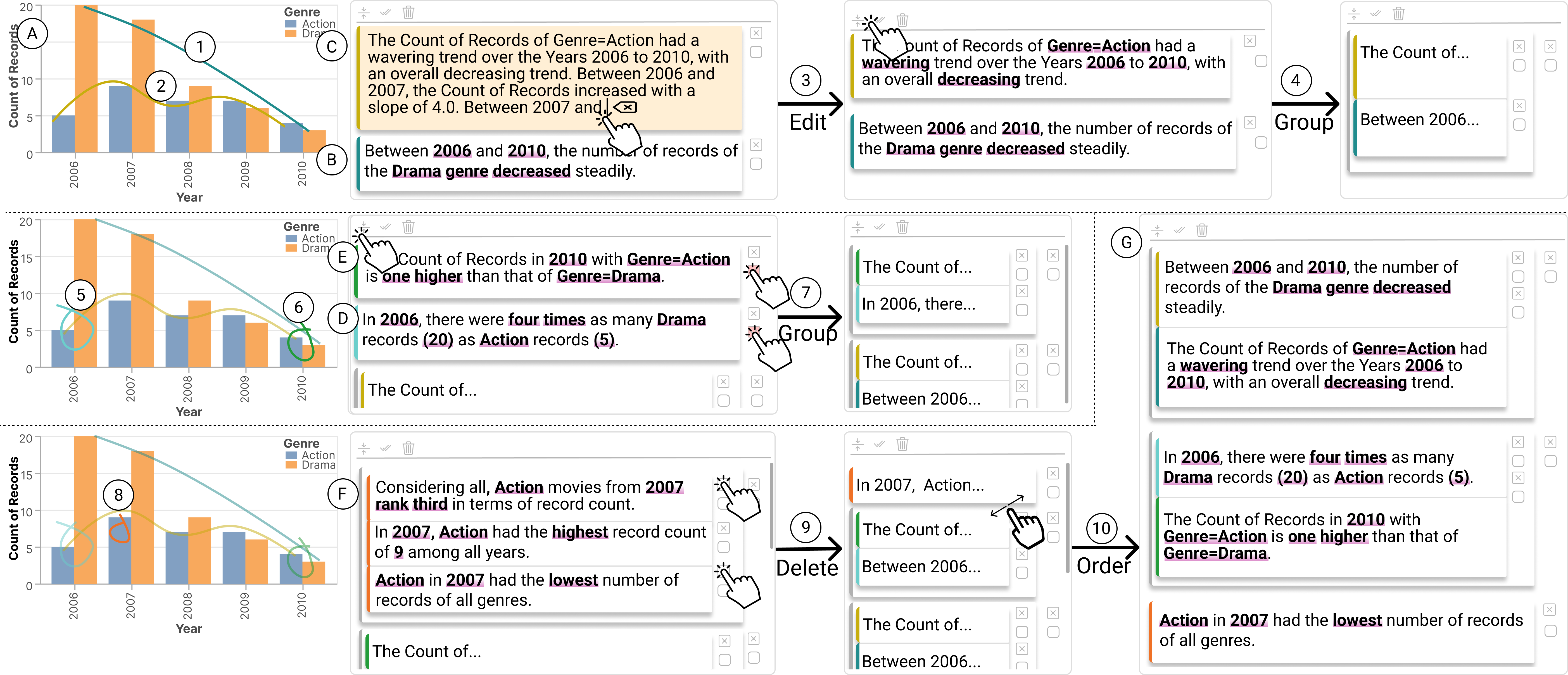}
    \caption{This figure illustrates a usage scenario in which Alice, an analyst, utilizes \system{} to document her findings during analyzing the movie market. (1-10) represent Alice's interactions with the chart, such as sketching, editing, and ordering. (A) corresponds to the chart that Alice created for her analysis, while (B-G) refer to the documentation results after the interaction. }
    \label{fig:scenario}
    \vspace{-1em}
\end{figure*}

In this section, we present a scenario to illustrate how \system{} assists users in documenting findings.
Suppose that Alice is a data analyst at an investment company.
She uses Jupyter Notebook to explore and analyze the movie market.
\Cref{fig:scenario} shows the finding documentation process of Alice, with further details provided below.
Alice is particularly interested in two movie genres: Action and Drama. 
To visualize the number of movies released in these genres between 2006 and 2010, Alice employs a grouped bar chart (\Cref{fig:scenario} (A)).

Upon examining the chart, Alice is first drawn to the decreasing trend of drama movies. 
She sketches a line following the trend of drama movies (\Cref{fig:scenario} (1)).
\system{} generates a documentation card with the description of this decreasing trend (\Cref{fig:scenario} (B)). 
Compared to drama movies, the number of released action movies appears more steady and wavering. 
Alice draws another line that resembles the trend (\Cref{fig:scenario} (2)).
The tool generates the second documentation card inserted at the top of the documentation panel (\Cref{fig:scenario} (C)).
It not only outlines the overall wavering and decreasing pattern but also specifies each time period  an increasing or decreasing trend appears (\Cref{fig:scenario} (C)).
Although accurate, Alice considers this level of detail excessive.
She clicks the documentation card to edit it and removes details about every small time interval (\Cref{fig:scenario} (3)).
Satisfied with the content, Alice thinks both the two documentation cards (\Cref{fig:scenario} (A)-(B)) are about trends.
Thus, she clicks the \textit{Group All} button at the top of the panel to group them (\Cref{fig:scenario} (4)).

Wanting to emphasize the changes in the differences between these two genres at the beginning (2006) and the end (2010) of the data range, Alice encircles the two bars in 2006 and the two bars in 2010 (\Cref{fig:scenario} (5)-(6)). 
She subsequently obtains two documentation cards about the comparison of the two genres. They together reveal a contrast that the count of drama movies in 2006 was four times greater than that of action movies (\Cref{fig:scenario} (D)), while in 2010, the count of drama movies was one less than that of action movies (\Cref{fig:scenario} (E)).
Alice groups these two documentation cards about the comparisons together by selecting the corresponding cards and clicking the \textit{Group} button at the top of the panel (\Cref{fig:scenario} (7)).

Furthermore, Alice highlights the number of action movies in 2007 by drawing a circle atop the corresponding blue bar 2007 (\Cref{fig:scenario} (8)). 
In response, \system{} offers three comparison options: comparing the selected bar to all other bars, to those representing action movies, and to those also from the year 2007.
It generates three documentation cards grouped together to form a larger card (\Cref{fig:scenario} (F)). 
Alice finds that the second one highlighting that in 2007 the number of action movies is largest compared to other years aligns with her intent.
She thus decides to delete the other two (\Cref{fig:scenario} (9)).

Finally, Alice feels satisfied with the content of these documentation cards in a hierarchical structure.
However, she finds that the cards could be reordered to form a better narrative flow. 
She drags and drops the documentation cards to order them (\Cref{fig:scenario} (10)), beginning with the general trends of the two genres, followed by comparing the genres at specific time points, and ending with highlighting the highest number of action movies in 2007.
This way, Alice effectively documents her findings for this chart with the help of \system{} (\Cref{fig:scenario} (G)).

\section{User Study}

To evaluate the usability and effectiveness of \system{}, we conducted a within-group user study with 12 participants.
They were asked to use \system{} and a baseline method with two datasets and give both qualitative and quantitative feedback.
We introduce the study setting in~\Cref{sec: study design} and then analyze the results obtained from the user study in~\Cref{sec: result analysis}.

\subsection{Study Setup}
\label{sec: study design}

\textbf{Participants.}
\cll{R2.13}
\yanna{Participants (P1-P12, 7 males and 5 females, aged 22 to 30, at an average age 25.5) were recruited through online advertisements on social networks and word of mouth.}
\yanna{They were} well-educated Ph.D. students of diverse backgrounds, including visualization, human-computer interaction, database, robotics, and software engineering.
When asked about their prior experience with Jupyter Notebook, Python, and data analysis, their self-rated average scores \yanna{were} 5.6, 5.9, and 5.3, respectively (1 = ``no experience'', 7 = ``highly experienced''). 
None of them \yanna{had} color blindness.
Each participant received compensation of US \$9/hour for completing the user study.

\begin{figure}[t!]
    \centering
    \includegraphics[width=1\linewidth]{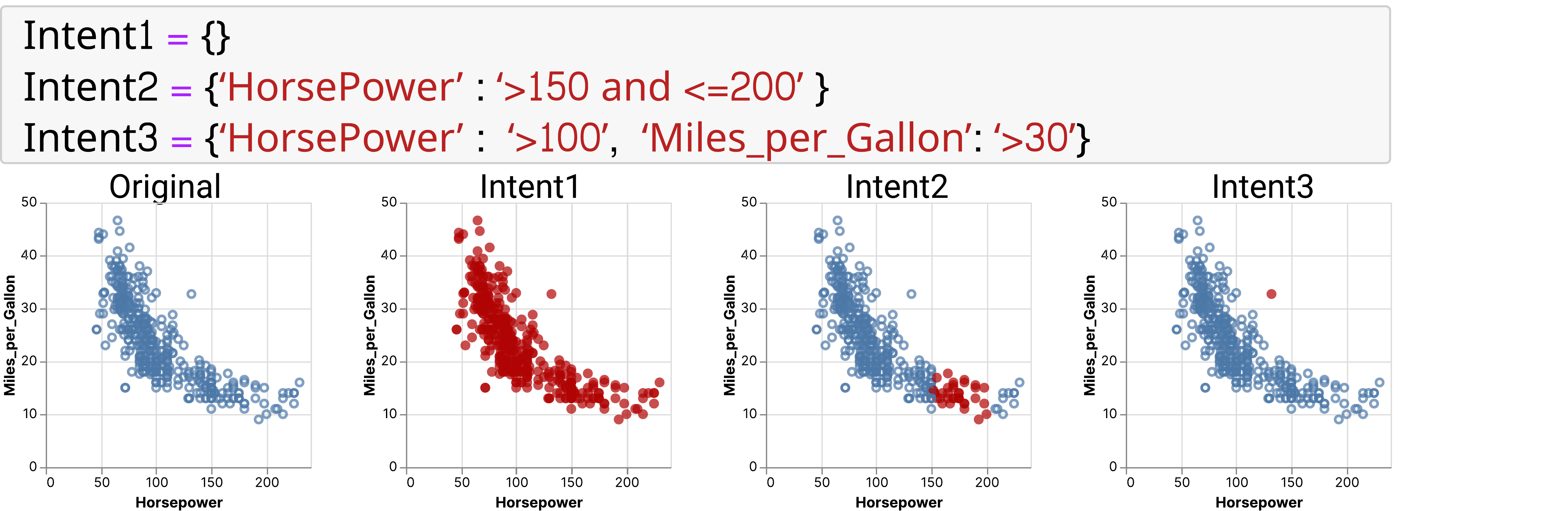}
    \caption{This figure shows \yanna{the baseline condition in the user study where users write code to specify their intent.} Specifically, \textit{Intent1} without any filter conditions selects all data items, \textit{Intent2} selects a data range using \textit{and}, while \textit{Intent3} selects a specific data item by limiting the values for two attributes. The selected data items are highlighted as red points.}
    \label{fig:intent}
    \vspace{-1em}
    
\end{figure}

\textbf{Baseline Method.} 
\cl{SR.1, R2.3, R4.2, R4.3}
\yanna{
\system{} enables users to express their intents by sketching.
The baseline approach required users to write code to express their intent in JSON format.
This code-based method, having been employed in prior research to specify user intent, such as Lux~\cite{lee2021lux} and Voyager2~\cite{Kanit2017voyager2}, offers comparable expressiveness to sketch.
}
\cl{R1.4, R3.3}
\yanna{
Despite the difference in how users express their intent in \system{} and the baseline, other aspects remain consistent; that is, selecting the same data of interest via either method will yield identical findings in the generated documentation.
Specifically, in \system{}, users sketch their intent atop the visualization directly. 
In the baseline approach, users create a new code cell to express their intent in JSON format,  which is then executed to generate the corresponding documentation of findings alongside the visualization. 
\Cref{fig:intent} showcases three examples of how users can write codes to specify their intent in baseline, which was also used as tutorial material in the user study.}

\textbf{Task.}
We designed an open-ended data analysis task that involved both exploring the data and documenting findings from data charts created by themselves during the exploration. 
We \yanna{chose} this design instead of directly giving participants charts for them to document for two reasons:
(1) we \yanna{attempted} to verify whether our tool can be naturally and smoothly integrated into an exploration flow;
(2) it was found that individuals would analyze the data charts with more intentions when the charts are created by themselves compared to when the charts are made by others~\cite{kim2019inking}. 
Participants were asked to explore a given dataset with charts and utilize the provided tools to document their findings, which they would later share with others.
Participants would finish the tasks with two tools in two different datasets.
The order of the tools and datasets was counterbalanced and randomly assigned to participants to reduce the learning effect.
We required that the participants document their findings using at least 4 charts \yanna{or within 30 minutes}.
During our own exploration and a pilot study with two other researchers out of this project, 
we \yanna{found} this number of charts \yanna{was} more likely to allow the tool to be fully explored without tiring people.

\textbf{Data.}
We selected the \textit{movies} and \textit{cars} datasets from the Vega dataset\footnote{\url{https://vega.github.io/vega-datasets/}} and \textit{house price prediction} dataset from the Kaggle dataset\footnote{\url{https://www.kaggle.com/competitions/house-prices-advanced-regression-techniques/data}}.
These datasets have been widely used to evaluate the effectiveness and usability of visualization tools~\cite{wang2023slide4n, wu2021multivision, wang2022documentation}. 
The \textit{cars} dataset was used only in the tutorial to familiarize participants with the tools. 
To make the task manageable in the user study, we reduced the dataset size to 10 attributes and approximately 500 rows.
All datasets contain the three common data types, \ie~temporal, categorical, and numerical.

\begin{figure*}[ht!]
    \centering
    \includegraphics[width=1\linewidth]{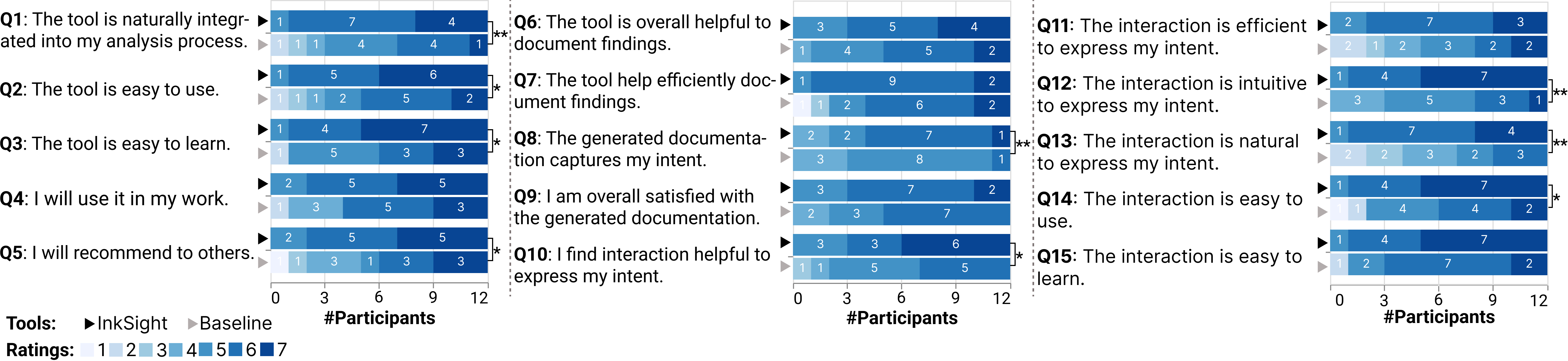}
    \caption{\yanna{This figure displays the user ratings for two tools: \system{} and the baseline, with \textbf{*} indicating $p<0.05$ and \textbf{**} indicating $p<0.01$. In general, \system{} outperforms the baseline across all aspects, with some aspects showing statistically significant differences. }}
    \label{fig:userstudy}
    \vspace{-1em}
\end{figure*}

\textbf{Procedure.}
The user study lasted about 1.5 hours and all of them were conducted through one-to-one offline meetings.
We started by introducing the procedure and asked for the participant's consent for recording the entire session.
Participants then completed a brief tutorial session to become familiar with \system{} or the \textit{baseline}. 
They first watched a short video about the components and functionalities and then tried them on a sample notebook with the \textit{cars} dataset. 
This tutorial session took around 10 minutes until participants felt familiar enough to use the tool.
Participants were then given a dataset with attribute descriptions and asked to explore data and document chart findings for approximately 30 minutes. 
After that, they were asked to briefly share their recorded findings.
Then, participants were asked to complete a \yanna{7-point Likert} questionnaire in a think-aloud protocol, where users were encouraged to justify their ratings. 
\yanna{Specifically, a rating of 1 means ``strongly disagree'', while a rating of 7 means ``strongly agree''.}
As shown in \Cref{fig:userstudy}, the questionnaires concerned three aspects, \ie~the usability and effectiveness of the tools (Q1-Q5 and Q6-Q7, respectively), the quality of the generated documentation (Q8-Q9), and the usability and effectiveness of the way to express intentions (Q10-Q15).
Before using the second tool, participants were given a five-minute break. 
The same procedure was conducted with the second tool. 
After the session, we conducted a post-interview to gather participants' opinions on the documentation quality, the pros and cons of the two tools to express intent, the preferred interaction method, and the pros and cons of \system{}.

\subsection{Result Analysis}
\label{sec: result analysis}
\cl{R1.3}
\yanna{During our user study, we observed that participants created 47 visualizations and documented 78 findings using the baseline. 
While using \system{}, they created 54 visualizations and documented 159 findings.
There was a statistically significant boost in the number of findings created with \system{} than the baseline ($p<0.01$). 
The number of visualizations and findings generated by each participant can be found in the supplementary material. 
In the following sections, we delve deeper into the quantitative and qualitative results of the user study.
}

\subsubsection{Quantitative results}
Next, we report the quantitative questionnaire results regarding three aspects: \ie~the usability and effectiveness of the tools (Q1-Q7), the quality of the generated documentation (Q8-Q9), and the usability and effectiveness of the way to express intentions (Q10-Q15).
\Cref{fig:userstudy} shows the quantitative results of our user study reflecting the average and variance of participants' ratings.
Specifically, we performed the Wilcoxon test to compare our tool \system{} and the baseline in all aspects.
We \yanna{denoted a statistically} significant difference with ``*'' for $p<0.05$, and ``**'' for $p<0.01$.
As shown in the results, \system{} \yanna{received} overall higher ratings than the baseline in all perspectives.

For the usability and effectiveness of the tools (Q1-Q7), \system{} \yanna{outperformed} the baseline \yanna{statistically} significantly in terms of usability (Q1-Q5), while being slightly better in effectiveness (Q6-Q7), as depicted in \Cref{fig:userstudy}.
Both \system{} and the baseline were recognized as being helpful and efficient for documenting findings, with all average ratings larger than 5.3 (Q6-Q7).
As a result, all participants expressed their willingness to use the tools in their work, with average ratings larger than 5.4 (Q4).
However, the difference in the ways of expressing intent led to \system{} outperforming the baseline \yanna{statistically} significantly in terms of its seamless integration into the data analysis workflow, ease of use and learning, and recommendation to others (with all $p<0.05$) (Q1-Q3 and Q5).

For the quality of generated documentation (Q8-Q9), participants overall recognized the generated documentation.
\system{} \yanna{statistically} significantly outperformed the baseline in capturing the users' intents (p < 0.01). 
However, no \yanna{statistically} significant differences were observed in the overall satisfaction with the generated documentation.
Since \system{} and the baseline \yanna{shared} the same computation modules, the difference in performance \yanna{was} likely due to the difference in the interaction method used to express intents.

For the way to express intent (Q10-Q15), the user study results revealed that the sketching interaction of \system{} \yanna{statistically} significantly outperformed the baseline interaction.
Specifically, sketching was more helpful (Q10), intuitive (Q12), natural (Q13), and easier to use (Q14) in expressing users' intent, with all of these aspects showing statistically significant differences (p < 0.05).   
As for efficiency (Q11) and ease to learn (Q15), there \yanna{was} no  \yanna{statistically} significant difference between the two interactions.
   
In conclusion, the results suggest that \system{} consistently \yanna{outperformed} the baseline across various aspects related to the overall tool, the quality of documentation, and interactions for expressing intent. 
Although some aspects did not show \yanna{statistically} significant differences, \system{} consistently achieved higher average ratings in the user study.

\subsubsection{Qualitative results}

This section presents participants' qualitative feedback from their justifications of questionnaire ratings and post-interview responses.


\cl{R2.10}
\yanna{\textbf{\system{} streamlines the documentation process and provides additional benefits, including enhancing data analysis and fostering effective collaboration.}}
All participants in the study appreciated \system{} for its ability to help them document findings. 
They recognized the tool's well-designed interactions and provided documentation effectively reducing their efforts in documenting findings. 
They specifically appreciated the seamless integration of \system{} into their analysis workflow, as it did not disrupt their data analysis process (\textbf{G3} and \textbf{G5}). 
For instance, P11 noted that ``\textit{\system{} with sketch aligns well with my analysis habits. When I record findings, I observe the points and trends and take notes on the charts directly.}''
Similarly, P3 mentioned that ``\textit{The code-free sketch makes \system{} seamlessly integrate into my data analysis workflow without any additional workload...I even did not realize when I was switching between exploring the data and recording findings.}''
Furthermore, participants recognized how the documentation of findings by \system{} can benefit collaboration and self-recall.
P1 mentioned that ``\textit{Placing the documentation next to the visualization instead of below it, which distinguished it from the analysis code, helps my future self and my cooperators to better locate the documentation.}''
P6 noted that ``\textit{Sharing my findings with my cooperators through \system{} allowed them to better understand what I have found, and they can easily add their own findings using the same method.}''

In addition to its benefits in documenting findings, participants highlighted other advantages brought by  \system{}, such as facilitating data analysis and fostering a sense of partnership. 
\system{} sometimes serves as a tool for users to obtain and inspect related data findings regarding a data subset of interest, which in turn helps some participants analyze data.
P5 mentioned that \textit{"\system{} helped me test my hypotheses on interesting data subsets quickly. I even did not need to write the code to analyze it"}.
Interestingly, \system{} can also give a feeling of being accompanied by serving as an AI partner, transforming the typically mundane documentation process into a more engaging experience. 
P12 said, ``\textit{Every time I draw a sketch, the tool responds to my input. It feels like I'm having a conversation with a partner. I appreciate this sense of companionship when documenting}''.

In conclusion, by integrating seamlessly with users' existing habits and offering a range of helpful features, \system{} has proved to be an effective tool for assisting in documenting findings during data analysis.


\textbf{\haotian{The sketch interaction in \system{} is considered an easy and flexible way to express intent by the majority of the participants.}}
10 out of 12 participants preferred the sketching method in \system{}, with the remaining two participants (P2 and P12) expressing their equal preferences for both methods. 
Participants found the code-free sketching to be more natural, intuitive, and flexible, allowing them to work directly on the chart with little workload (\textbf{G2}).
In particular, 
the sketch-based interaction enabled users to express their intent by drawing any shapes to highlight what they see in the chart, while the coding method demands additional mental efforts to identify attribute names and values carefully.
For example, P8 mentioned that ``\textit{When sketching, I just highlight the area that my eye catches. While for coding, I need to further process what I see into codes.}''
P5 noted that ``\textit{When coding my intent, I just feel like I work as a compiler to translate my intent to a grammar, which is really unnatural and unfriendly.}''
Additionally, 
when talking about the baseline method, participants said that they often made mistakes when typing out their intent in a rigorous programming format. 
Sometimes the tedious process could cause participants to give up using the tool, as P1 pointed out ``\textit{I would prefer to spend my time documenting findings directly rather than writing extra codes first for expressing intent.}''

Despite the advantages of sketch interaction, we find a personal preference for the baseline method.
P2 and P12, who rated their Python programming experience highly (7 out of 7), stated that they enjoyed programming and felt comfortable and more familiar with using programming to accomplish tasks. 
However, P12 added that ``\textit{The convenience of the sketch is more beneficial to open exploration since I can draw any shapes to see the results with a low burden. While coding is more suitable for focused analysis since I need to think about it carefully.}''

\textbf{\haotian{The quality of documentation generated by \system{} is thought highly by the participants.}}
In the user study, participants expressed overall satisfaction with the generated documentation, finding it helpful to reduce their efforts in documenting findings (\textbf{G1}). 
They appreciated the natural language used in the documentation, the efficiency gained from using it, and the fact that they could easily edit and customize it (\textbf{G4}).
P5 commented that ``\textit{The generated ones is what I needed, and most of the time, I can share to others directly.}''
Some participants acknowledged that the data facts covered many common and useful aspects, providing a comprehensive understanding of the data.
P6 remarked ``\textit{The tool provides almost all the basic data facts that I need for data analysis.}''

However, the generated documentation did not always fully capture participants' intent.
For example, P10 would like to highlight two peak points in a line by drawing an open-path sketch following the whole line.
Unexpectedly, \system{} only described the general trend without explicitly mentioning the two peaks.
Nonetheless, unexpected data facts could sometimes have positive effects.
Three participants (P4, P8, P12) realized that these facts beyond their expectations could provide them with a new perspective of data analysis and inspire their next-step exploration.
For instance,
P4 noted that ``\textit{Unexpected information can help me avoid overlooking important details.}''

Overall, while the generated documentation was well-rated, there is still room for improvement in terms of comprehensiveness.

\section{Discussion}
In this section, we discuss the lessons learned, the limitation of this study, and future research directions. 


\subsection{Design Lessons}
\textbf{Human-AI collaboration in \system{}.}
\system{} applies a Human-AI collaboration workflow for documenting chart findings.
When using \system{}, humans express their intent through sketching first.
AI then generates initial drafts of documentation, and humans finally refine the documentation.
All participants appreciate the assistance of AI in reducing manual work in this tedious task.
Previous human-AI collaboration tools in computational notebooks~\cite{li2023notable, wang2022documentation} mainly focus on defining the roles of humans and AI for more efficient and effective task assignments.
Unlike them, we target interaction designs for facilitating the communication between humans and AI, which is an important challenge in facilitating human-AI collaboration\yanna{\cite{amershi2019guidelines, li2023ai}}.
The user study results suggest that our tool design can help AI capture participants' intent with little input from them.
Participants further provide valuable insights into future improvements in human-AI communication. 
Some participants desired a way to further specify their needs on the basis of the original input of intent when AI generates documentation that has redundant or irrelevant information.
In other words, they would like AI to revise the documentation based on updated intent
instead of editing the documentation themselves.
Furthermore, several participants appreciate personalized AI assistance, such as generating documentation aligned with their individual habits and styles. 
We will keep working on facilitating AI's understanding of human intent in a more iterative and personalized manner.

\textbf{Enabling user to express intents through multi-modal interaction methods.}
All participants in the user study appreciated the sketch interaction for expressing their intent, praising its intuitive and engaging nature. 
The sketching method also offered a higher degree of freedom, enabling users to draw any shape of interest without the constraints imposed by coding. 
However, users may accidentally select the data items incorrectly when selecting points distant or surrounded by undesired ones.
While the code-based approach was found to be more accurate, it came at the cost of a laborious process of reading the chart in detail to obtain attribute names and values and writing lengthy codes for filtering. 
Moreover, the code-based approach is more restrictive and demanding in terms of chart interpretation. 
Each interaction method has its own advantages and disadvantages. 
As such, one potential direction for future work is to support multi-modal interactions \yanna{(\eg~\cite{walny2012understanding})} simultaneously, maximizing the benefits of each method while minimizing their respective drawbacks. 
This would enable users to choose the most suitable interaction method based on their personal preferences, skills, and the specific task at hand.

\textbf{Documenting findings in computational notebooks.}
\haotian{In our research, we have some observations that may help deepen the understanding of documenting findings in computational notebooks.}
First, our study participants mentioned that documenting findings is essential, as they usually had some initial ideas when examining a chart that either inspired their next-step exploration or those they wanted to share with others. 
Due to the tedious process of documentation, they often prioritized data exploration over spending time on explanation. 
As a result, they sometimes forgot the rationale behind their analysis or the findings they wanted to share, leading to additional time spent revisiting charts to jog their memory.
Thus, documenting findings promptly during data exploration can help record their findings when the memory is fresh and save time for future reference.
Second, participants appreciated the placement of the documentation panel to the right of the visualization. 
This layout offers several benefits.
It enables users to locate corresponding documentation swiftly. 
Additionally, the horizontal placement of the documentation separates it from the vertically arranged code analysis, effectively distinguishing between data analysis and data exploration, facilitating users' focus on each aspect of their work.
Third, it is noteworthy that participants tended not to edit the documentation during the data analysis process. 
Instead, they briefly scanned the generated findings and simply deleted unwanted ones, leaving detailed documentation revision after completing data exploration. 
This observation indicates the importance of striking a balance between documentation and data analysis, ensuring that the documentation process does not impede the exploration flow.

\subsection{Limitation and Future Work}

\cl{R2.11}
\yanna{
\textbf{Enhancing the quality of documentation created by \system{}.}
Firstly, 
it will be helpful to support documenting findings based on multiple code cells and related charts.
Currently, \system{} focuses on applying sketch interaction to allow users to document findings based on one code cell and the associated chart. 
However, participants (P4, P8, P10, P12) in our user study expressed interest in enhancing documentation that spans multiple code cells and corresponding charts, \ie~integrating previous codes or prior finding documentation as context.
Such an enhancement would align with the interconnected nature of findings often observed during exploratory data analysis, where analysts frequently base subsequent explorations on previous findings~\cite{lee2019fallacy}. 
In the future, it will be interesting to enhance \system{} by integrating techniques like understanding code histories~\cite{will2022solas} and chart relationships~\cite{lin2023dashboard} for more contextualized documentation based on the entire exploration workflow.}
\cl{R1.1, \ R3.2}
\yanna{Secondly, there is a need for \system{} to support more types of charts and data facts.  
Currently, \system{} is limited to six chart types and eight types of data facts (see \Cref{sec:method}). 
However, participants expressed the need for more chart types (P4, P8, P10, P12), like heatmaps, and more diverse types of data facts (P2, P5, P9, P10), such as local minima and maxima.
Future work should focus on extending the coverage of supported chart types and data facts to enhance \system{}'s utility.}  
\cl{R4.6}
\yanna{Lastly, the documentation generated by \system{} can be more natural.
Currently, though we employ GPT-3.5 to organize and polish the description of data facts, some generated documentation can look template-based and not intuitive,
such as the expression "Genre=Action" in \Cref{fig:scenario} (C) and (E). 
To improve the documentation quality, it is possible to explore the usage of more advanced natural language processing models such as GPT-4~\cite{openai2023gpt4} or fine-tune the prompt in the future.
}

\yanna{\textbf{Improving the interaction design of \system{}.}
}
\cl{SR.2, R2.2}
\yanna{
Firstly, enhancing the interaction design of \system{} to minimize user effort in documenting findings can be an improvement avenue.
Presently, \system{} generated all data facts related to the data subset identified by users, allowing users to edit out unwanted ones.
While this strategy could save user effort in expressing intent, it might result in information overload when there are too many findings for users to consume and select from.
Another approach is to ask users to put more effort into specifying their intents. For example, Intentable~\cite{choi2022intentable} asks users to click intent types (e.g., ``overview'', ``trend'') and data items of interest to generate more focused results.
Therefore, it remains an open question of how to strike a balance between users' efforts in specifying their intent and their efforts in editing the automatically generated results.
Additional user studies could help understand user preferences and behaviours regarding these two different intent inference strategies for generating documentation, ultimately informing iterative interface improvements.
}\cl{R2.1}\yanna{Secondly, enhancing the interaction with sketch in \system{} offers a promising direction for improvement. 
Currently, to refine a sketch, users have to delete it, which limits users’ ability to refine their intent.
Future improvement of \system{} could consider integrating additional interactive features, such as dragging, resizing, and erasing sketches~\cite{browne2011sketchvis, chung2022talebrush}, thereby providing users with greater flexibility in adjusting their intent.
}

\section{Conclusion}

Computational notebooks provide data analysts with a convenient approach to exploratory data analysis by combining code cells and outputs within a single document.
However, effectively documenting chart findings remains a challenge due to its time-consuming and tedious nature.
To address the challenge, this study explores a natural and convenient way to reduce the workload of document chart findings.
Specifically, we develop \system{}, a computational notebook plugin, to document chart findings according to users' intent leveraging sketch interaction.
It allows users to sketch their interested data items and creates corresponding documentation with language generation models.
\system{} was considered effective by the participants in our user study.
They agreed that \system{} offered an effortless experience of finding documentation through natural and engaging sketch interaction.
Furthermore, they pointed out future improvements of \system{}, such as capturing intent more accurately and generating documentation that considers context and semantics.
In the future, we hope to continue the research on facilitating documenting chart findings.
\yanna{Potential directions enhancing the documentation generation and refining the interface and interaction design of \system{}.}
\cl{R4.4}
\yanna{Additionally, we hope to conduct long-term and real-world user studies with more diverse participants.
\haotian{Through these studies, we can deepen our understanding of users' workflow and experience of documenting and communicating data analysis results with \system{} and other tools (\eg Notable~\cite{li2023notable}) in computational notebooks.}}

\bibliographystyle{abbrv-doi-hyperref}

\bibliography{template}


%

\appendix 







\end{document}